\documentclass{article}
\usepackage{spconf,amsmath,graphicx}

\usepackage{color}
 
\usepackage{bm}
\usepackage{amsmath,amssymb,amsfonts,graphicx,amsthm,mathtools,nicefrac} 
\usepackage{algorithm, algorithmicx,algpseudocode}
\usepackage[numbers]{natbib}
 
\usepackage{subfigure}

\DeclareMathOperator{\rank}{\mathrm{rank}}

\DeclareMathOperator{\dist}{\mathrm{dist}}

\def\la {\left\langle}
\def\ra {\right\rangle} 
\newcommand{\matsnorm}[2]{\left\| #1\right\|_{{#2}}}

\newcommand{\fronorm}[1]{\ensuremath{\matsnorm{#1}{\footnotesize{\mathsf{F}}}}}
\newcommand{\opnorm}[1]{\ensuremath{\matsnorm{#1}{}}}
\newcommand{\twoinf}[1]{\ensuremath{\matsnorm{#1}{\footnotesize{\mbox{2,$\infty$}}}}}

\newcommand{\twonorm}[1]{\ensuremath{\matsnorm{#1}{\footnotesize{2}}}}

\newcommand{\bfm}[1]{\bm{#1}}

\newcommand{\E}[2][]{\mathbb{E}_{#1} \left\{ #2 \rule{0mm}{3mm}\right\}}

\def\va{\bfm a}     
\def\vb{\bfm b}   \def\mB{\bfm B}  
     \def\C{\mathbb{C}}
     
\def\ve{\bfm e}     
    
\def\vg{\bfm g}     
\def\vh{\bfm h}     
   \def\mI{\bfm I}

   \def\mL{\bfm L}  
   \def\mM{\bfm M}

   \def\mP{\bfm P}  
   \def\mQ{\bfm Q}  
   \def\mR{\bfm R}  \def\R{\mathbb{R}}

   \def\mU{\bfm U}  
   \def\mV{\bfm V}  
     
\def\vx{\bfm x}   \def\mX{\bfm X}  
\def\vy{\bfm y}

\def\calA{{\cal  A}}

\def\calD{{\cal  D}}

\def\calH{{\cal  H}} 
\def\calI{{\cal  I}}

\def\calM{{\cal  M}} 
 
\def\calO{{\cal  O}} 
\def\calP{{\cal  P}}

\newcommand{\bfsym}[1]{\bm{#1}}

             \def\bSigma{\bfsym \Sigma}

              \def\hbSigma{\widehat{\bfsym \Sigma}}

\def \tran {\mathsf{T}}
\def \tranH{\mathsf{H}}

\def \hU {\widehat{\mU}}
\def \hV {\widehat{\mV}}
\def \hL {\widehat{\mL}}
\def \hR {\widehat{\mR}}

\newtheorem{theorem}{Theorem}[section]
\newtheorem{assumption}{Assumption}[section]
\newtheorem{remark}{Remark}[section]

\numberwithin{equation}{section}

\title{Blind Super-resolution via Projected Gradient Descent }
 
\name{Sihan Mao, ~Jinchi Chen\thanks{This work is partially supported by National Science Foundation of China under Grant No. 12001108.}}
\address{School of Data Science, Fudan University, Shanghai, China}
  
\begin{document}
 
\maketitle
\begin{abstract}

Blind super-resolution can be cast as low rank matrix recovery problem by exploiting the inherent simplicity of the signal. In this paper, we develop a simple yet efficient non-convex method for this problem based on the low rank structure of the vectorized Hankel matrix associated with 
the target matrix. 
Theoretical guarantees have been established under the similar conditions as convex approaches. Numerical experiments are also conducted to demonstrate its performance.
%apply the low rank structure of the vectorized Hankel matrix corresponding to the target matrix and propose a computational efficient method for this problem. 
%We show that a simple projected gradient descent method started from an initial guess exhibits good behavior and is more efficient than start-of-the-art convex optimization methods,  yet still requiring similar exact recovery conditions. The theoretical findings are demonstrated by numerical experiments.
\end{abstract}
\section{Introduction}
\label{sec:intro}
Blind super-resolution of point sources is the problem of simultaneously estimating locations and amplitudes of point sources and  point spread functions from low-resolution measurements. Such problem arises in various applications, including single-molecule imaging \cite{quirin2012optimal}, medical imaging \cite{qu2015accelerated}, multi-user communication system \cite{luo2006low} %quirin2012optimal} 
and so on. 

Under certain subspace assumption and applying the lifting technique, blind super-resolution can be cast as a matrix recovery problem. Recent works in \cite{chi2016guaranteed, yang2016super, li2019atomic, suliman2021mathematical, chen2020vectorized} exploit the intrinsic structures of data matrix and propose different convex relaxation methods for such problem. Theoretical guarantees for these methods have been established. However, due to the limitations of convex relaxation, all of these methods do not scale well to the high dimensional setting. 

In contrast to convex relaxation, a non-convex recovery method is proposed in this paper based on the Vectorized Hankel Lift \cite{chen2020vectorized} framework. More precisely, 
harnessing low-rank structure of vectorized Hankel matrix corresponding to the signal in terms of the Burer-Monteiro factorization, we develop a projected gradient descent algorithm, named PGD-VHL, to directly recover the low rank factors. We show that such a simple algorithm possesses a remarkable reconstruction ability. Moreover, our algorithm started from an initial guess converges linearly to the target matrix under the similar sample complexity as convex approaches.

The rest of this paper is organized as follows. We begin with the problem formulation and describe the proposed algorithm in Section \ref{sec: problem formulation}. Section \ref{sec: main results} provides a convergence analysis of PGD-VHL.  Numerical experiments to illustrate the performance of PGD-VHL are provided in Section \ref{sec: numerical}. Section \ref{sec: conclusion} concludes this paper.

\section{Algorithm}
\label{sec: problem formulation}
\subsection{Problem formulation}
The point source signal can be represented as a superposition of $r$ spikes 
\begin{align*}
	x(t)  = \sum_{k=1}^{r}d_k \delta(t- \tau_k),
\end{align*}
where $d_k$ and $\tau_k$ are the amplitude and location of $k$-th point source respectively. Let $\{g_k(t)\}_{k=1}^r$ be the unknown point spread functions. The observation is a convolution between $x(t)$ and $\{g_k(t)\}_{k=1}^r$,
\begin{align*}
	y(t ) = \sum_{k=1}^{r}d_k \delta(t-\tau_k) \ast g_k(t) = \sum_{k=1}^{r}d_k g_k(t-\tau_k).
\end{align*}
After taking Fourier transform and sampling, we obtain the measurements as 
\begin{align}
\label{eq: measurement}
y[j] = \sum_{k=1}^{r}d_k e^{-2\pi \imath \tau_k\cdot j}\cdot \hat{g}_k[j] ~~\text{ for } j=0,\cdots, n-1.
\end{align}
Let $\vg_k = \begin{bmatrix}
	\hat{g}_k[0]&\cdots &\hat{g}_k[n-1]
\end{bmatrix}^\tran$ be a vector corresponding to the $k$-th unknown point spread
function. The goal is to estimate $\{d_k, \tau_k\}_{k=1}^r$ as well as $\{\vg_k\}_{k=1}^r$ from \eqref{eq: measurement}. Since the number of unknowns is larger than $n$, this problem is an ill-posed problem without any additional assumptions. Following the same route as that in \cite{chi2016guaranteed, yang2016super, li2019atomic, chen2020vectorized}, we assume $\{\vg_k\}_{k=1}^r$ belong to a known subspace spanned by the columns of $\mB\in\C^{n\times s}$, i.e., 
\begin{align*}
	\vg_k = \mB\vh_k.
\end{align*}
Then under the subspace assumption and applying the lifting technique \cite{ahmed2013blind}, the measurements \eqref{eq: measurement} can be rewritten as a linear observations of $\mX^\natural: = \sum_{k=1}^{r}d_k \vh_k \va_{\tau_k}^\tran\in\C^{s\times n}$:
\begin{align}
	\label{eq: linear measurements 1}
	y_j = \la \vb_j \ve_j^\tran, \mX^\natural \ra\quad \text{for}\quad j=0,\cdots, n-1,
\end{align}
where $\vb_j$ is the $j$-th row of $\mB$, $\ve_j$ is the $(j+1)$-th standard basis of $\R^n$, and $\va_{\tau}\in\C^n$ is the vector defined as 
\begin{align*}
	\va_{\tau} = \begin{bmatrix}
		1 & e^{-2\pi\imath \tau\cdot 1} &\cdots &e^{-2\pi\imath \tau\cdot (n-1)}
	\end{bmatrix}^\tran.
\end{align*} 
The measurement model \eqref{eq: linear measurements 1} can be rewritten succinctly as 
\begin{align}
	\label{eq: linear measurements 2}
	\vy = \calA(\mX^\natural),
\end{align}
where $\calA:\C^{s\times n}\rightarrow \C^n$ is the linear operator. Therefore, blind super-resolution can be cast as the problem of recovering the data matrix  $\mX^\natural$ from its linear measurements \eqref{eq: linear measurements 2}. 

Let $\calH$ be the vectorized Hankel lifting operator %\cite{chen2020vectorized} 
which maps a matrix $\mX\in\C^{s\times n}$ into an $sn_1\times n_2$ matrix, 
\begin{align*}
	\calH(\mX) = \begin{bmatrix}
		\vx_0 & \vx_1 &\cdots & \vx_{n_2-1}\\
		\vx_1 & \vx_2 &\cdots & \vx_{n_2}\\
		\vdots& \vdots &\ddots &\vdots\\
		\vx_{n_1-1}&\vx_{n_1}&\cdots &\vx_{n-1}\\
	\end{bmatrix}\in\C^{sn_1\times n_2},
\end{align*}
where $\vx_i$ is the $(i+1)$-th column of $\mX$ and $n_1+n_2 = n+1$. It is shown that $\calH(\mX^\natural)$ is a rank-$r$ matrix \cite{chen2020vectorized} and thus the matrix $\calH(\mX^\natural)$ admits low rank structure when $r \ll \min(sn_1, n_2)$. It is natural to recover $\mX$ by solving the constrained least squares problem
% explain
\begin{align*}
    \min_{\mX}~\frac{1}{2}\twonorm{\vy - \calA(\mX)}^2 ~\text{ s.t. } \rank(\calH(\mX)) = r.
\end{align*}

To introduce our algorithm,  we assume that $\mX^\natural$ is $\mu_1$-incoherent which is defined as below.
\begin{assumption}
\label{assumption 1}
	Let $\calH(\mX^\natural) = \mU\bSigma\mV^\tranH$ be the singular value decomposition of $\calH(\mX^\natural) $, where $\mU\in\C^{sn_1\times r}, \bSigma\in\R^{r\times r}$ and $\mV\in\C^{n_2\times r}$. Denote $\mU^\tranH = \begin{bmatrix}
		\mU_0^\tranH &\cdots &\mU_{n_1-1}^\tranH
	\end{bmatrix}^\tranH$, where $\mU_\ell= \mU[\ell s + 1 : (\ell+1) s, :]$ is the $\ell$-th block of $\mU$ for $\ell=0,\cdots n_1-1$. The matrix $\mX^\natural$ is $\mu_1$-incoherence if $\mU$ and $\mV$  obey that
	\begin{align*}
		\max_{0\leq \ell\leq n_1-1} \fronorm{\mU_\ell}^2 \leq \frac{\mu_1 r}{n} \text{ and }\max_{0\leq j \leq n_2-1} \twonorm{\ve_j^\tran\mV}^2 \leq \frac{\mu_1 r}{n}
	\end{align*}
	for some positive constant $\mu_1 $.
\end{assumption}
\begin{remark}
Assumption \ref{assumption 1} is the same as the one made in \cite{candes2009exact,zhang2018multichannel} for low rank matrix recovery and is used in blind super-resolution \cite{chen2020vectorized}. It has been established that Assumption \ref{assumption 1} is obeyed when the minimum wrap-up distance between the locations of point sources is greater than about $1/n$. 
\end{remark}
Letting $\mL^\natural = \mU\bSigma^{1/2}$ and $\mR^\natural = \mV\bSigma^{1/2}$, we have
\begin{align*}
	\max_{0\leq \ell\leq n_1-1} \fronorm{\mL_\ell^\natural} \leq \sqrt{\frac{\mu_1 r\sigma_1}{n}} \text{ and } \twoinf{\mR^\natural} \leq \sqrt{\frac{\mu_1 r\sigma_1}{n}},
\end{align*}
where $\sigma_1 = \opnorm{\calH(\mX^\natural)}$.  Since that target data matrix $\mX^\natural$ is $\mu_1$-incoherence and the low rank structure of the vectorized Hankel matrix can be promoted by $\calH(\mX^\natural) = \mL^\natural {\mR^\natural}^\tranH$, it is natural to recover the low rank factors of the ground truth matrix $\calH(\mX^\natural)$ by solving an optimization problem in the form of 
\begin{align}
	\label{eq: optimization 1}
	\min_{\mM\in\calM }~\bigg\{ f(\mM):=&\frac{1}{2}\twonorm{\vy - \calA\calH^\dagger(\mL\mR^\tranH)}^2 \notag\\
	&+ \frac{1}{2}\fronorm{ \left(\calI - \calH\calH^\dagger\right)(\mL\mR^\tranH)}^2 \notag \\
	&+ \frac{1}{16}\fronorm{\mL^\tranH\mL - \mR^\tranH\mR}^2\bigg\},
\end{align}
where $\mM = \begin{bmatrix}
	\mL^\tranH &\mR^\tranH
\end{bmatrix}^\tranH\in\C^{(sn_1+n_2)\times r}$, $\calH^\dagger$ is the Moore-Penrose pseudoinverse of $\calH$ obeying that $\calH^\dagger\calH = \calI$, the second term in objective guarantees that $\mL\mR^\tranH$ admits vectorized Hankel structure. Last term penalizes the mismatch between $\mL$ and $\mR$, which is widely used in rectangular low rank matrix recovery \cite{tu2016low, zheng2016convergence,chi2019nonconvex}. The  convex feasible set $\calM$ is defined as follows
\begin{align}
\label{eq: convex set}
	\calM = \bigg\{ \begin{bmatrix}
		\mL\\
		\mR\\
	\end{bmatrix}~:~ \max_{0\leq \ell\leq n_1-1} \fronorm{\mL_\ell} &\leq \sqrt{\frac{\mu r\sigma }{n}},\notag \\
 \twoinf{\mR} &\leq  \sqrt{\frac{\mu r\sigma}{n} }\bigg\}, 
\end{align}
where $\mL_\ell$ is the $\ell$-th block of $\mL$, $\mu$ and $\sigma$ be two absolute constants such that $\mu \geq \mu_1$ and $\sigma \geq \sigma_1$.

\subsection{Projected gradient descent method}
\label{sec: pgd}
Inspired by \cite{cai2018spectral}, we design a projected gradient descent method for the problem \eqref{eq: optimization 1}, which is summarized in  Algorithm \ref{alg: PGD-VHL}. 
\begin{algorithm}[h]
	\caption{PGD-VHL}
	\label{alg: PGD-VHL}
	\begin{algorithmic}
		\State Input: $\calA, \vy, n, s, r$
        \State Initialization:
		\State $~\quad n_1 = n/2, \quad n_2 = n+1-n_1$
		\State $~\quad \hU_0 \hbSigma_0 {\hV_0}^\tranH=  \calP_r\calH\calA^\ast(\vy) $
	    \State $~\quad \hL_0 =\hU_0 {\hbSigma_0}^{1/2},\quad \hR_0 = \hV_0 {\hbSigma_0}^{1/2}$
		\State $~\quad (\mL_0, \mR_0) = \calP_{\calM}((\hL_0, \hR_0))$
		\State $~\quad \mM_0 = \begin{bmatrix}\mL_0^\tranH &\mR^\tranH_0 \end{bmatrix}^\tranH$

		\While{ not convergence}
		\State $\mM_{t+1} = \calP_{\calM}\left( \mM_t - \eta \nabla f(\mM_t) \right)$.
		\EndWhile
	\end{algorithmic}
	%\hspace*{0.02in} {\bf Output:}
	%$\widehat\mZ^{t}$ in the last iteration, $\vy^t = \calG^H(\mZ_1^t{\mZ_2^t}^H)$, and $\mZ^t = \mD^{-1}\calG^H(\mZ_1^t{\mZ_2^t}^H)$.
\end{algorithm}
The initialization involves two steps: (1) computes the best rank $r$ approximation of $\calH(\calA^\ast(\vy))$ via one step hard thresholding $\calP_r(\cdot)$, where $\calA^\ast$ is the adjoint of $\calA$ ;%and defined as $\calA^\ast(\vy) = \sum_{i=0}^{n-1}y_i\vb_i \ve_i^\tran$; 
(2) projects the low rank factors of best rank-$r$ approximated matrix onto the set $\calM$.  Given a matrix {\small $\mM = \begin{bmatrix}\mL^\tranH & \mR^\tranH\end{bmatrix}^\tranH$}, the projection onto $\calM$, denoted by {\small $\begin{bmatrix}\widehat{\mL}^\tranH & \widehat{\mR}^\tranH\end{bmatrix}^\tranH$}, has a closed form solution:
\begin{align*}
	[\widehat{\mL}]_\ell = \begin{cases}
		\mL_\ell&\text{ if }\fronorm{\mL_\ell} \leq \sqrt{\frac{\mu r\sigma}{n}}\\
		\frac{1}{\fronorm{\mL_\ell}}\mL_\ell\cdot \sqrt{\frac{\mu r\sigma}{n}} &\text{ o.w.}
	\end{cases}
\end{align*}
for $0 \leq \ell \leq n_1-1$ and 
\begin{align*}
	\ve_j^\tran\widehat{\mR} = \begin{cases}
		\ve_j^\tran \mR &\text{ if } \twonorm{\ve_j^\tran\mR} \leq \sqrt{\frac{\mu r\sigma}{n}}\\
		\frac{\ve_j^\tran\mR}{\twonorm{\ve_j^\tran\mR}}\cdot \sqrt{\frac{\mu r\sigma}{n}} &\text{ o.w.}
	\end{cases}
\end{align*}
for $0\leq j \leq n_2-1$.  Let $\mM_t$ be the current estimator. The algorithm updates $\mM_t$ along gradient descent direction $-\nabla f(\mM_t) $ with step size $\eta$, followed by projection onto the set $\calM$.
The gradient of $f(\mM)$ is computed with respect to Wirtinger calculus  given by %$\nabla f = \begin{bmatrix}{(\nabla_{\mL} f)}^\tranH, {(\nabla_{\mR} f)}^\tranH \end{bmatrix}^\tranH$, 
$\nabla f =\begin{bmatrix}\nabla^\tranH_{\mL} f& \nabla^\tranH_{\mR} f \end{bmatrix}^\tranH  $
where 
%\begin{align*}
%	\nabla f = \begin{bmatrix}
%		\nabla_{\mL} f\\
%		\nabla_{\mR} f\\
%	\end{bmatrix},
%\end{align*}
%where 
\begin{align*}
	\nabla_{\mL} f= &\left(\calH\calD^{-2}\calA^\ast \left(\calA\calH^\dagger(\mL\mR^\tranH) - \vy \right) \right)\mR \\
	&+ \left( \left(\calI - \calH\calH^\dagger\right)(\mL\mR^\tranH) \right)\mR +\frac{1}{4}\mL\left(\mL^\tranH \mL - \mR^\tranH\mR\right),\\
	\nabla_{\mR} f = & \left(\calH\calD^{-2}\calA^\ast \left(\calA\calH^\dagger(\mL\mR^\tranH) - \vy \right) \right)^\tranH\mL \\
	&+ \left( \left(\calI - \calH\calH^\dagger\right)(\mL\mR^\tranH) \right)^\tranH\mL+\frac{1}{4}\mR\left(\mR^\tranH\mR - \mL^\tranH \mL \right).
\end{align*}
To obtain the computational cost of $\nabla f$, we first introduce some notations. Let $\calH_{v}$ be the Hankel operator which maps a vector $\vx\in\C^{1\times n}$ into an $n_1\times n_2$ matrix,
\begin{align*}
	\calH_{v}(\vx) = \begin{bmatrix}
		x_1 &\cdots &x_{n_2}\\
		\vdots &\ddots &\vdots\\
		x_{n_1}&\cdots &x_n\\
	\end{bmatrix},
\end{align*}
where $x_i$ is the $i$-th entry of $\vx$.  The adjoint of $\calH_{v}$, denoted by $\calH_{v}^\ast$, is a linear mapping from $n_1\times n_2$ to $1\times n$. It is known that the computational complexity of both $\calH_{v}^\ast(\mL_\ell\mR^\tranH)$ and $\left(\calH_v(\vx)\right)\mR$ is $\calO(rn\log n)$ flops \cite{cai2019fast}. Moreover, the authors in \cite{chen2020vectorized} show that $\calH(\mX) = \mP\widetilde{\calH}(\mX)$, where $\widetilde{\calH}(\mX)$ is a matrix constructed by stacking all $\{\calH_{v}(\ve_\ell^\tran\mX)\}_{\ell=1}^s$ on top of one another, and $\mP$ is a permutation matrix. Therefore we can compute $\calH^\dagger(\mL\mR^\tranH)$ and $\calH(\mX)\mR$ by using $\calO(srn\log n)$ flops. Thus the implementation of our algorithm is very efficient and the main computational complexity in each step is $\calO(sr^2n + srn\log (n))$.

\section{Main results}
\label{sec: main results}
In this section, we provide an analysis of PGD-VHL under a random subspace model. 

\begin{assumption} 
\label{assumption 2}
The column vectors $\{\vb_j\}_{j=0}^{n-1}$ of $\mB$ are independently and identically drawn from a distribution $F$ which satisfies the following conditions
	\begin{align}
		\label{eq: isotropy property}
		\E{\vb_j\vb_j^\tranH} &= \mI_s, \quad j=0,\cdots, n-1,\\
		\label{eq:incoherence}
		\max_{0\leq \ell\leq s-1}  | \vb_j[\ell]|^2 &\leq \mu_0, \quad j=0,\cdots, n-1.
	\end{align}
\end{assumption}
\begin{remark}
Assumption \ref{assumption 2} is a standard assumption in blind super-resolution \cite{chi2016guaranteed, yang2016super, li2019atomic, suliman2018blind,chen2020vectorized}, and holds with $\mu_0 = 1$ by many common random ensembles, for instance, the components of $\vb$ are Rademacher random variables taking the values $\pm 1$ with equal probability or $\vb$ is uniformly sampled from the rows of a Discrete Fourier Transform
(DFT) matrix.
\end{remark}
Now we present the main result of the paper.
%Before presenting the main theorem, it should be mentioned that for the tuning parameter $\sigma$ in \eqref{eq: convex set}, we will show in the later that $\sigma$ can be chosen as $\sigma = \opnorm{\textcolor{red}{\hZ_0}}/(1-\varepsilon)$ for some $0<\varepsilon<1$ and $\sigma\geq \sigma_1$ holds with high probability. The main result for the theoretical guarantee of PGD-VHL is presented as follows: 
\begin{theorem}
\label{main result}
Let $\mu \geq \mu_1$ and $\sigma = \sigma_1({\widehat\bSigma}_0)/(1-\varepsilon)$ for $0 \leq \varepsilon \leq 1/3$. Let $\eta\leq \frac{\sigma_r}{4500(\mu_0\mu s r)^2\sigma^2}$ and {\small$\mM^\natural = \begin{bmatrix}{\mL^\natural}^\tranH\quad{\mR^\natural}^\tranH\end{bmatrix}^\tranH$}. Suppose $\mX^\natural$ obeys the Assumption \ref{assumption 1} and the subspace $\mB$ satisfies the Assumption \ref{assumption 2}. If 
\begin{align*}
n\geq c_0 \varepsilon^{-2}\mu_0^2\mu s^2r^2 \kappa^2\log^2(sn),
\end{align*}
with probability at least $1-c_1 (sn)^{-c_2}$, the sequence $\{\mM_t\}$ returned by Algorithm 1 satisfies
	\begin{align}
		\label{ineq: convergence rate}
		\dist^2(\mM_t,\mM^\natural)\leq (1-\eta\sigma_r)^t\cdot \frac{\varepsilon^2\sigma_r}{\mu_0 s},
	\end{align}
	where $c_0, c_1, c_2$ are absolute constants, $\sigma_r = \sigma_r(\calH(\mX^\natural))$, $\kappa$ is the condition number of $\calH(\mX^\natural)$, and the distance $\dist(\mM,\mM^\natural)$ is defined as
	\begin{align*}
	   \dist(\mM,\mM^\natural) = \min_{\mQ\mQ^\tran = \mQ^\tran\mQ = \mI_r} \fronorm{\mM - \mM^\natural \mQ}.
	\end{align*}
\end{theorem}
\begin{remark}
% sample complexity
The detailed proof of Theorem \ref{main result} is provided in \cite{mao2021blind}.
Compared with the sample complexity established in \cite{chen2020vectorized} for the nuclear norm minimization method, which is $n\gtrsim \mu_0 \mu_1 \cdot sr\log^4(sn)$, Theorem \ref{main result} implies that PGD-VHL is sub-optimal in terms of $s$ and $r$. We suspect that it is merely an artifact of our proof. 

% linear convergence computation complexity
\end{remark}
\begin{remark}
Theorem \ref{main result} implies that PGD-VHL converges to $\mM^\natural$ with a linear rate. Therefore, after $T=\calO((\mu_0\mu sr\kappa)^2\log(1/\epsilon))$ iterations, we have $\dist^2(\mM_T, \mM^\natural) \leq \epsilon\cdot \dist^2(\mM_0, \mM^\natural) $. Given the iterates $\mM_T$ returned by PGD-VHL, we can estimate $\mX_T$ by $\calH^\dagger(\mL_T \mR_T^\tranH)$. 
\end{remark}
\begin{remark}
Once the data matrix $\mX^\natural$ is recovered, the locations $\{\tau_k\}_{k=1}^r$ can be computed from $\mX^\natural$ by MUSIC algorithm and the weights $\{d_k, \vh_k\}_{k=1}^r$ can be estimated by solving an overdetermined linear system \cite{chen2020vectorized}.
\end{remark}

\section{Numerical simulations}
\label{sec: numerical}
In this section, we provide numerical results to illustrate the
performance of PGD-VHL.
The locations $\{\tau_k\}$ of the point source signal is randomly generated from $[0,1)$ and the amplitudes $\{d_k\}$ are selected to be $d_k = (1+10^{c_k})e^{-\imath \phi_k}$, where $c_k$ is uniformly sampled from $[0,1]$ and $\phi_k$ is uniformly sampled from $[0,2\pi)$. The coefficients $\{\vh_k\}_{k=1}^r$ are i.i.d. sampled from standard Gaussian with normalization. %The noiseless measurement is obtained by \eqref{eq: linear measurements 1}, where 
The columns of %the subspace matrix 
$\mB$ are uniformly sampled from the DFT matrix. The stepsize of PGD-VHL is chosen via backtracking line search and PGD-VHL  will be terminated if $\twonorm{y-\calA(\mX_t)}\leq 10^{-5}$ is met or a maximum number of iterations is reached. We repeat 20 random trials and record the probability of successful recovery in our tests. A trial is declared to be successful if $\fronorm{\mX_t - \mX^\natural}/\fronorm{\mX^\natural} \leq 10^{-3}$. 

\begin{figure}[htb]

% 1 vhl
\begin{minipage}[b]{.48\linewidth}
  \centering
  \centerline{\includegraphics[width=4.0cm]{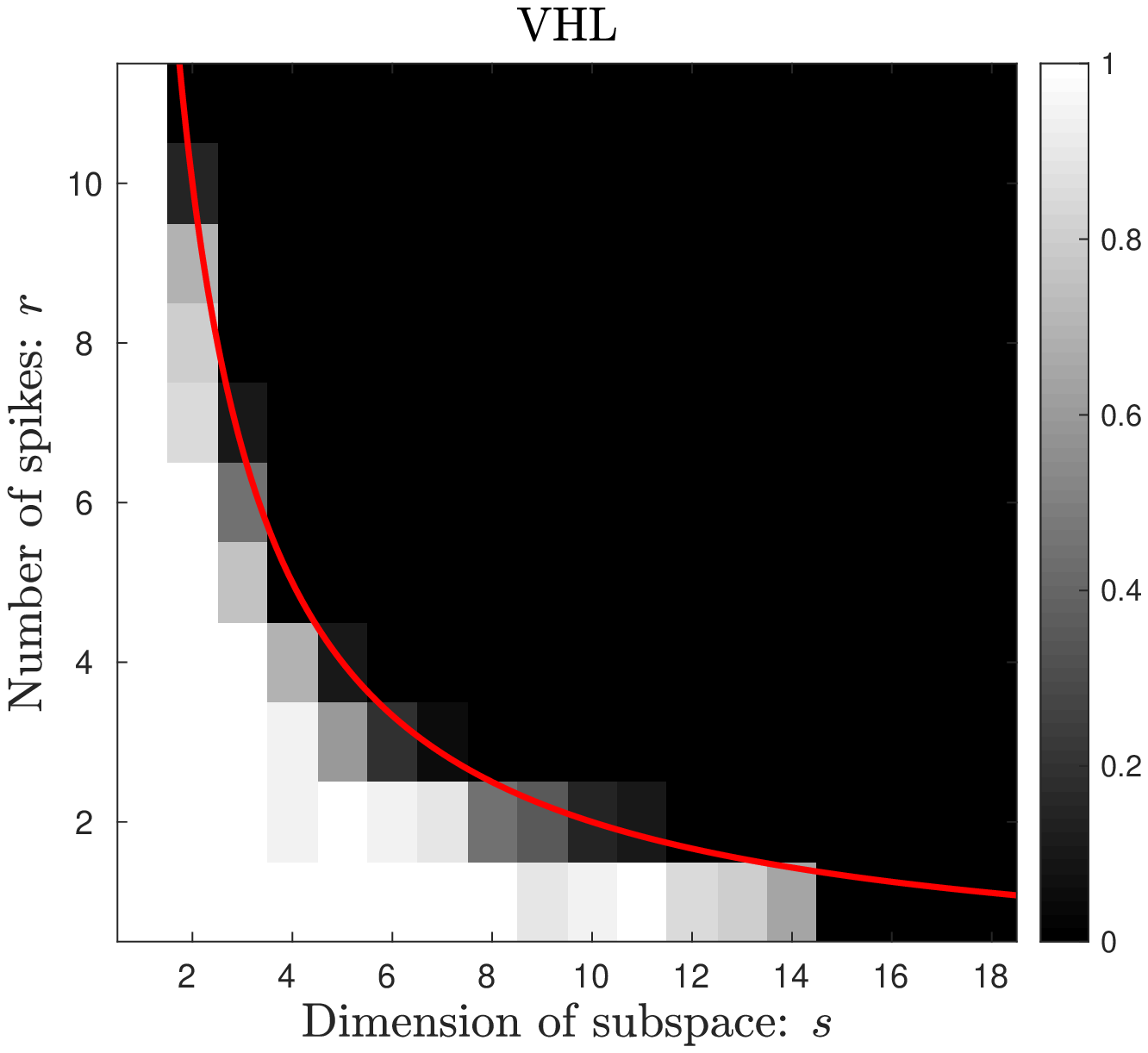}}
%  \vspace{1.5cm}
  \centerline{(a)}\medskip
\end{minipage}
\hfill
\begin{minipage}[b]{0.48\linewidth}
  \centering
  \centerline{\includegraphics[width=4.0cm]{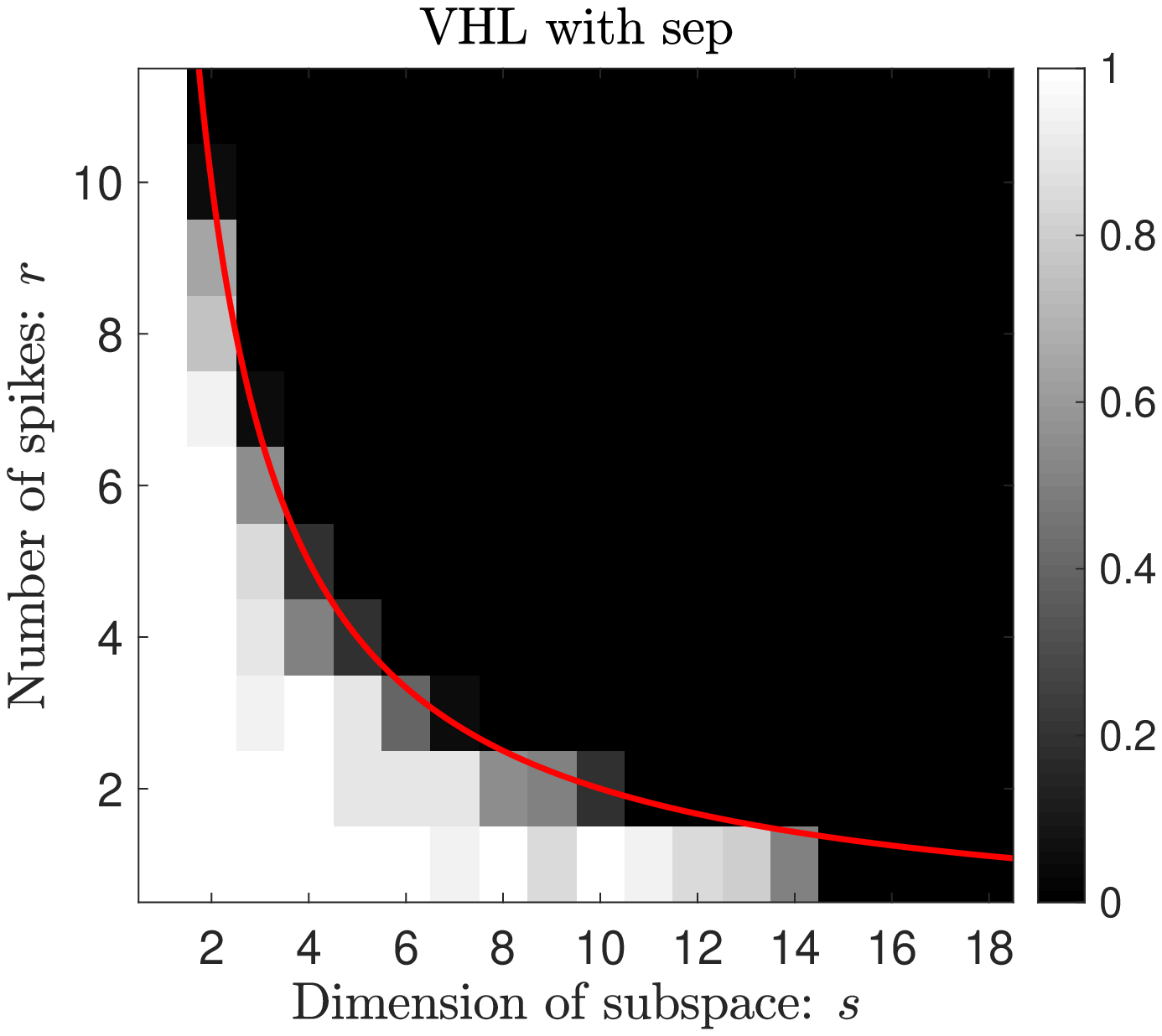}}
%  \vspace{1.5cm}
  \centerline{(b)}\medskip
\end{minipage}
%

% 2 anm
\begin{minipage}[b]{.48\linewidth}
  \centering
  \centerline{\includegraphics[width=4.0cm]{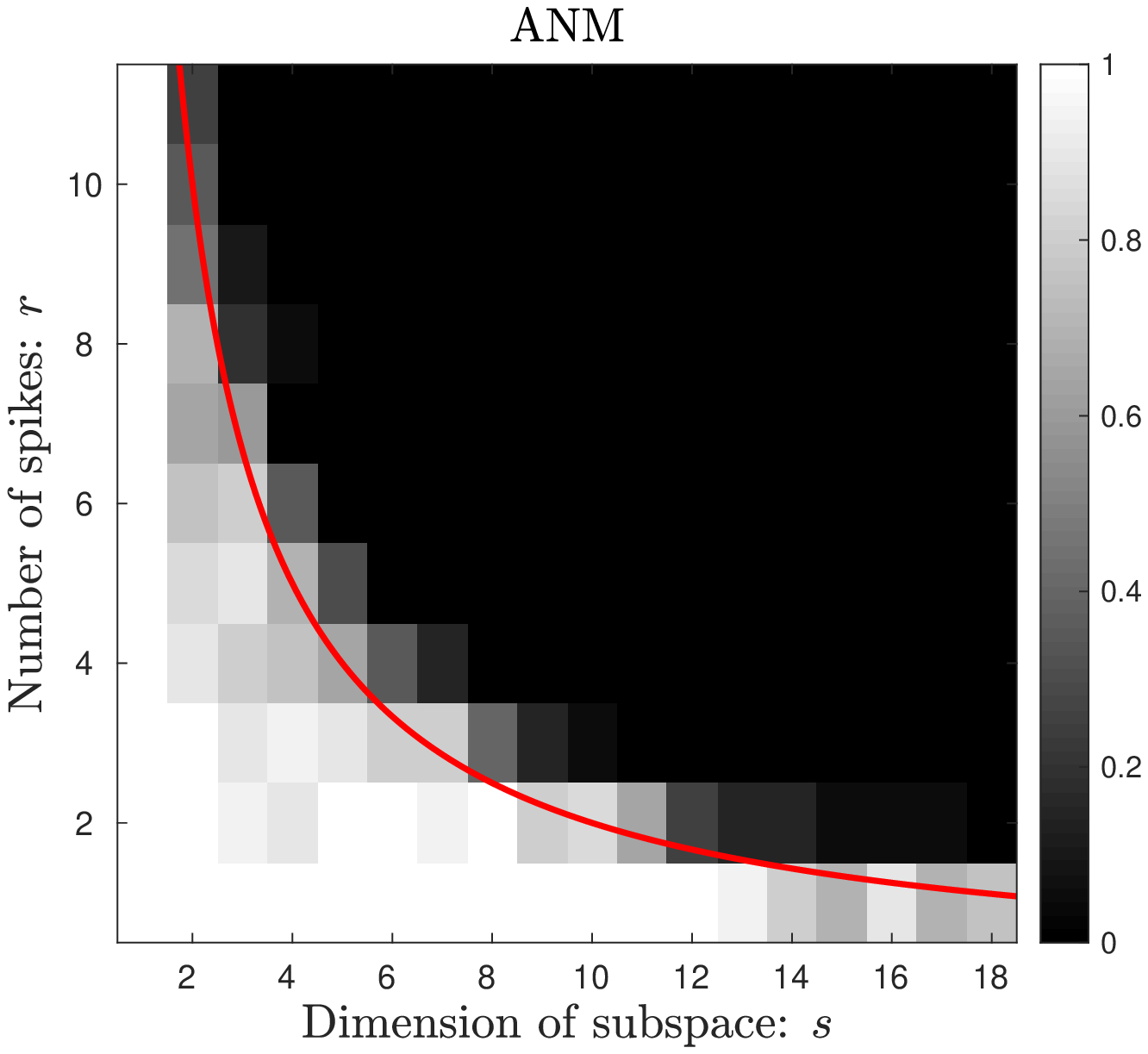}}
%  \vspace{1.5cm}
  \centerline{(c)}\medskip
\end{minipage}
\hfill
\begin{minipage}[b]{0.48\linewidth}
  \centering
  \centerline{\includegraphics[width=4.0cm]{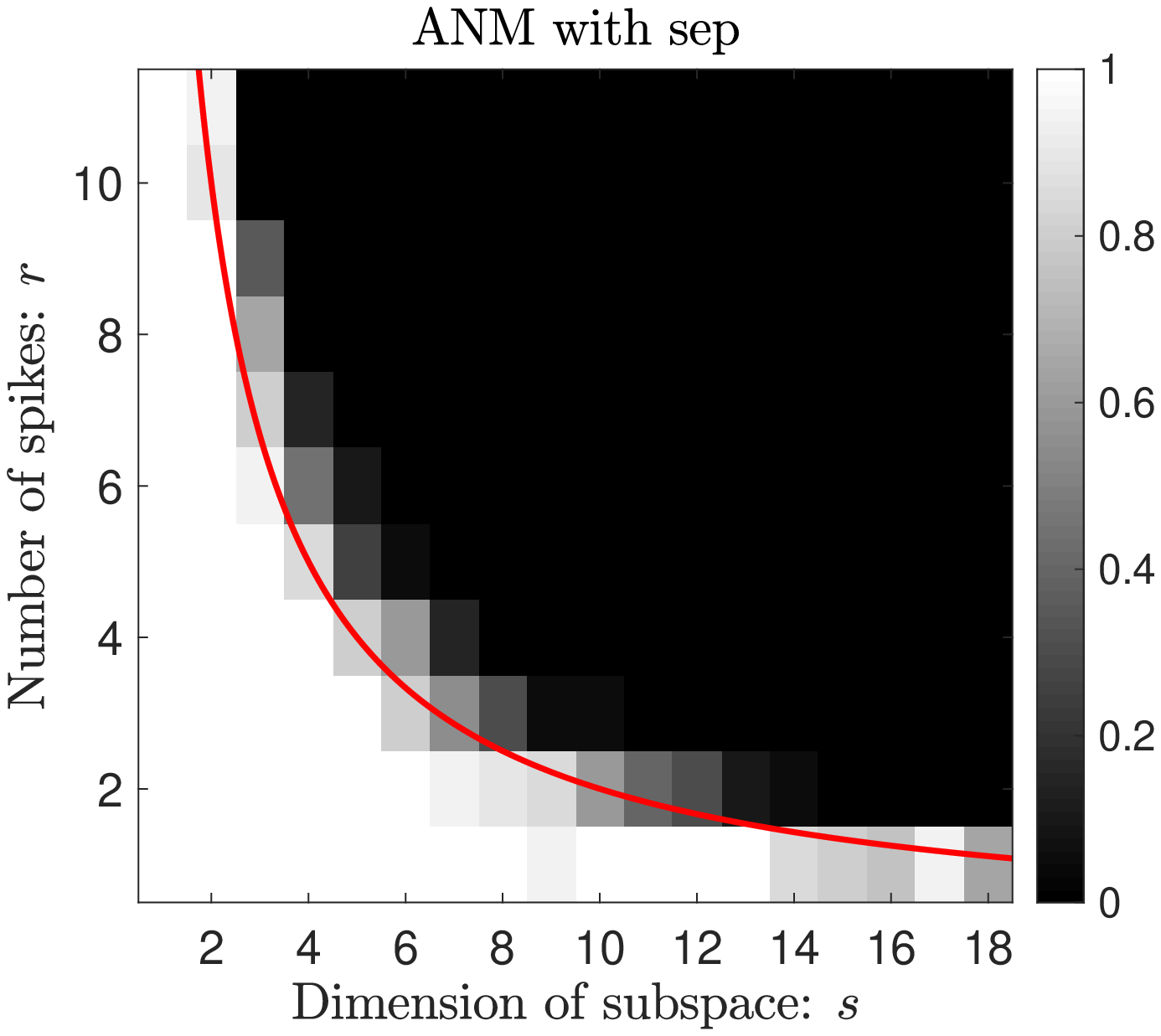}}
%  \vspace{1.5cm}
  \centerline{(d)}\medskip
\end{minipage}
%

% 3 pgd
\begin{minipage}[b]{.48\linewidth}
  \centering
  \centerline{\includegraphics[width=4.0cm]{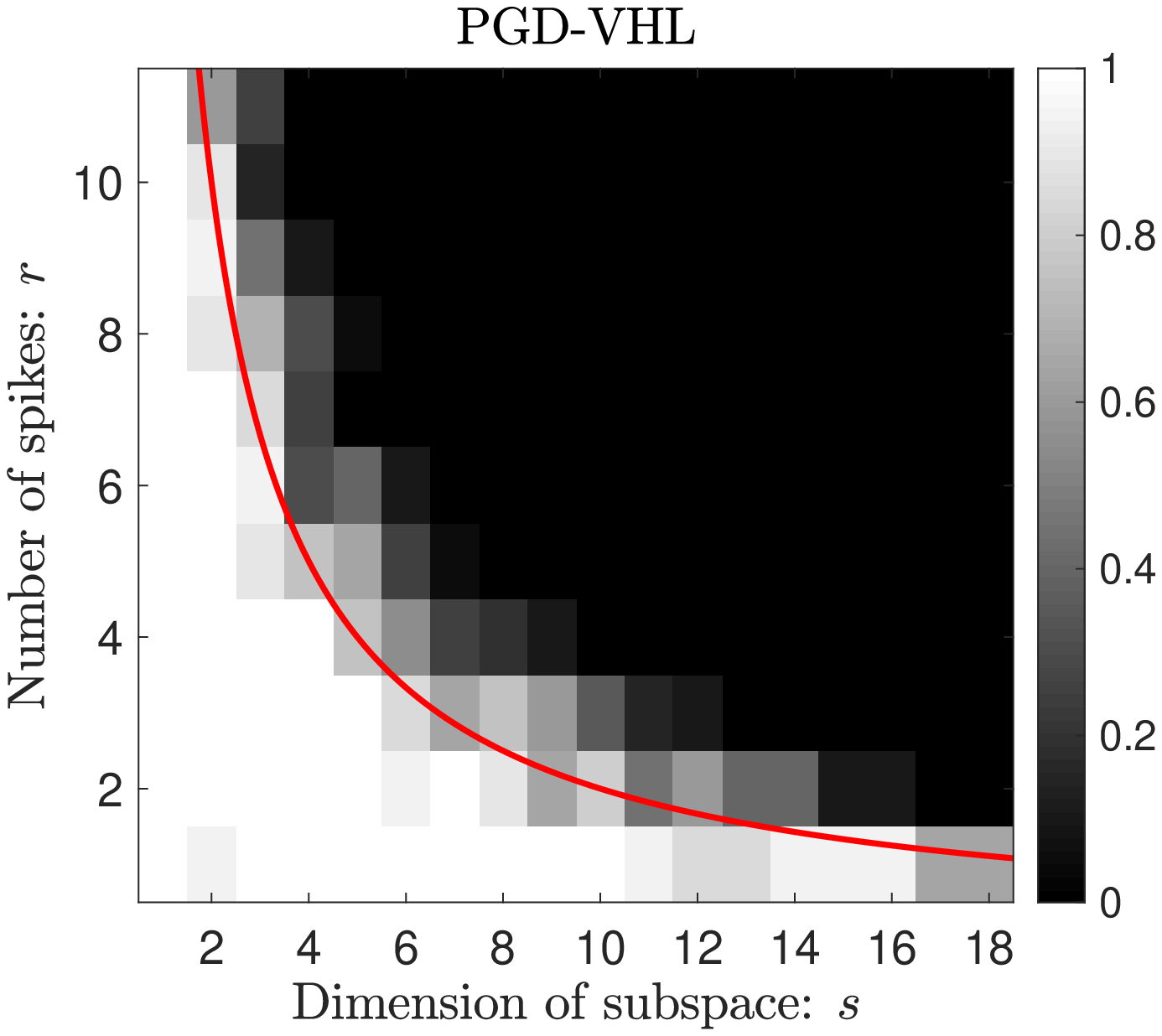}}
%  \vspace{1.5cm}
  \centerline{(e)}\medskip
\end{minipage}
\hfill
\begin{minipage}[b]{0.48\linewidth}
  \centering
  \centerline{\includegraphics[width=4.0cm]{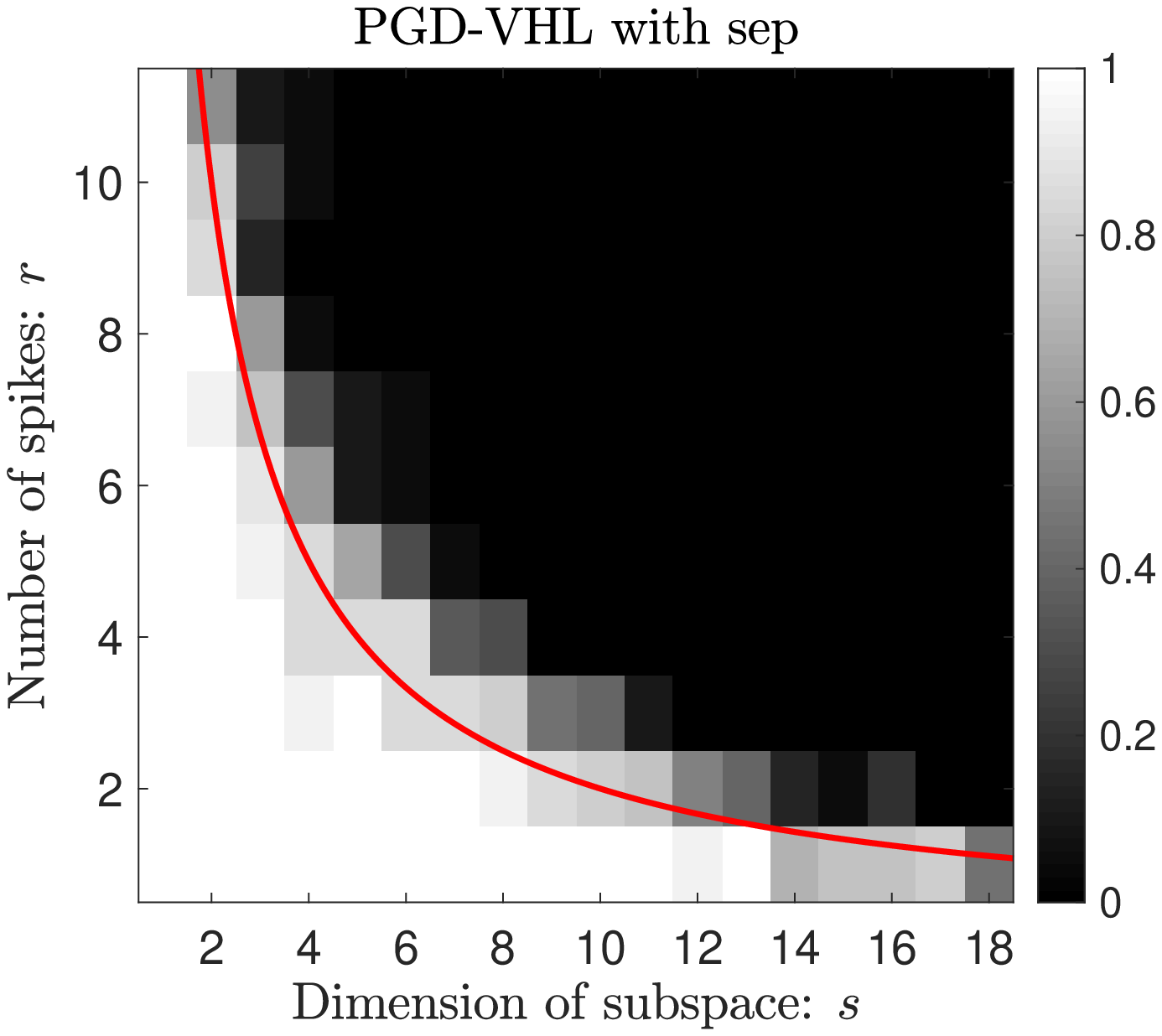}}
%  \vspace{1.5cm}
  \centerline{(f)}\medskip
\end{minipage}

\caption{The phase transitions of VHL (a,b), ANM (c,d) and PGD-VHL (e,f) with (Left) or without (Right) imposing the separation condition.
Here we fix $n=64$. The red curve plots the hyperbola curve $rs = 20$.}
\label{fig:phase transition 1}
\end{figure}

The first experiment studies the recovery ability of PGD-VHL through the framework of phase transition and we compare it with two convex recovery methods: VHL \cite{chen2020vectorized} and ANM \cite{yang2016super}. Both VHL and ANM are solved by CVX \cite{cvx}. 
The tests are conducted with $n=64$ and the varied $s$ and $r$. Figure \ref{fig:phase transition 1}(a), \ref{fig:phase transition 1}(c) and \ref{fig:phase transition 1}(e) show that phase transitions of VHL, ANM and PGD-VHL when the locations of point sources are randomly generated%without imposing the separation condition
, and Figure \ref{fig:phase transition 1}(b), \ref{fig:phase transition 1}(d) and \ref{fig:phase transition 1}(f) illustrate the phase transitions of VHL, ANM and PGD-VHL when the separation condition $\Delta:=\min_{j\neq k}|\tau_j -\tau_k| \geq 1/n $ is imposed. It can be seen that PGD-VHL is less sensitive to the separation condition than ANM and has a higher phase transition curve than VHL. 

%In Figure~\ref{fig:phase transition 1}, the phase transition curves for VHL, ANM and PGD-VHL are presented. It can be seen that PGD-VHL has a higher phase transition curve than that of VHL. 
%The top three subfigures plot the phase transitions when there is no separation condition imposed. The other three subfigures below present the phase transitions under the separation condition $\Delta:= \min\limits_{k\neq j}\lab\tau_k - \tau_j\rab\geq \frac{1}{n}$. 

%It can be observed that ANM has the highest phase transition when the separation of frequencies is satisfied, but degrades severely without separation condition. The phase transition of PGD-VHL with separation is slightly higher than that with no separation. The performance of VHL seems to be unaffected to the separation.    

In the second experiment, we study the phase transition of PGD-VHL when one of $r$ and $s$ is ﬁxed. Figure 2(a) indicates a approximately linear relationship between $s$ and $n$ for the successful recovery when the number of point sources is ﬁxed to be $r = 4$. The same linear relationship between $r$ and $n$ can be observed when the dimension of the subspace is ﬁxed to be $s = 4$, see Figure 2(b). Therefore there exists a gap between our theory and empirical observation and we leave it as future work.

\begin{figure}[htb]

	\begin{minipage}[b]{.48\linewidth}
		\centering
		\centerline{\includegraphics[width=4cm]{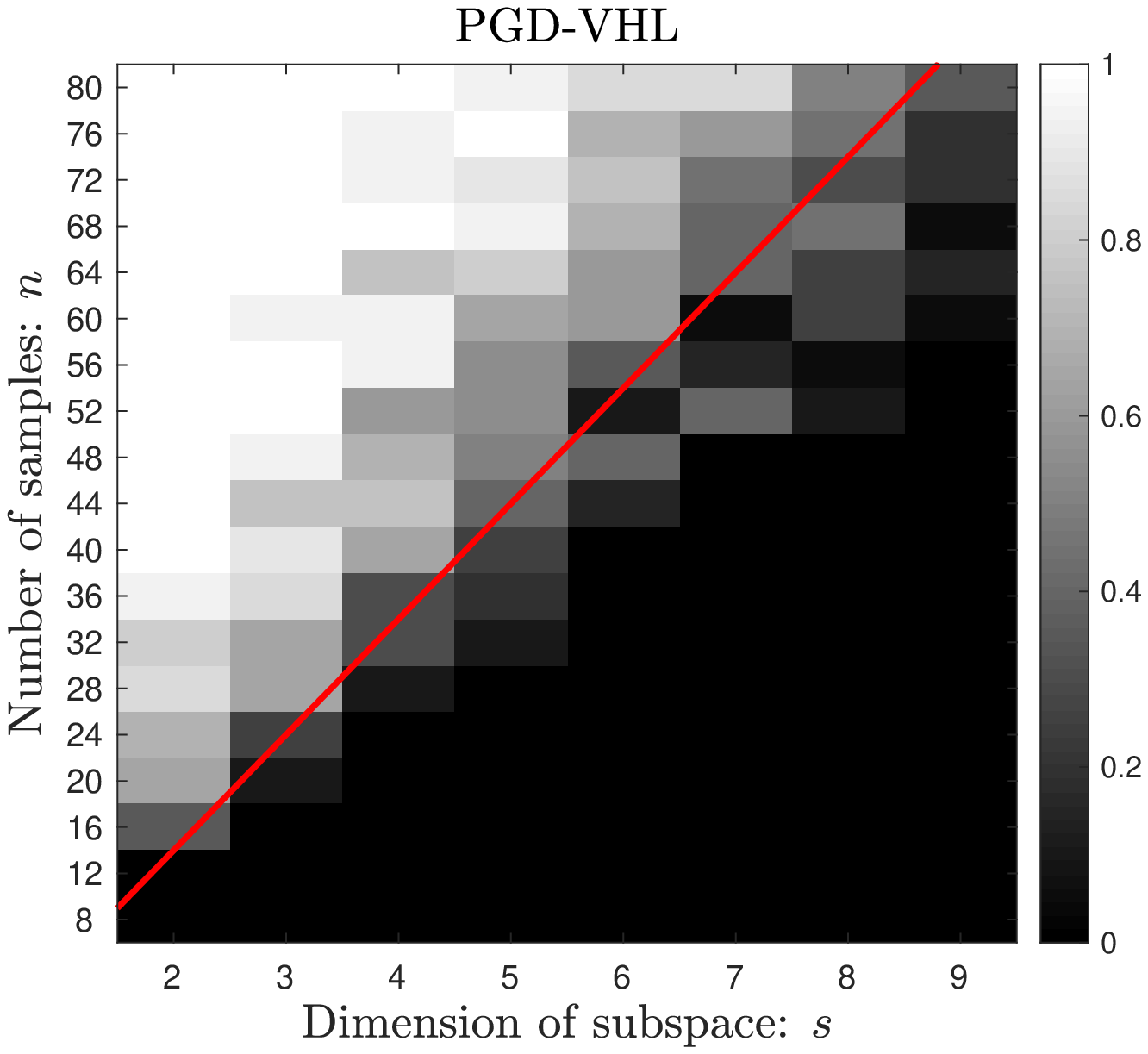}
	}
		%  \vspace{1.5cm}
		\centerline{(a)}\medskip
	\end{minipage}
	\hfill
	\begin{minipage}[b]{.48\linewidth}
		\centering
		\centerline{\includegraphics[width=4cm]{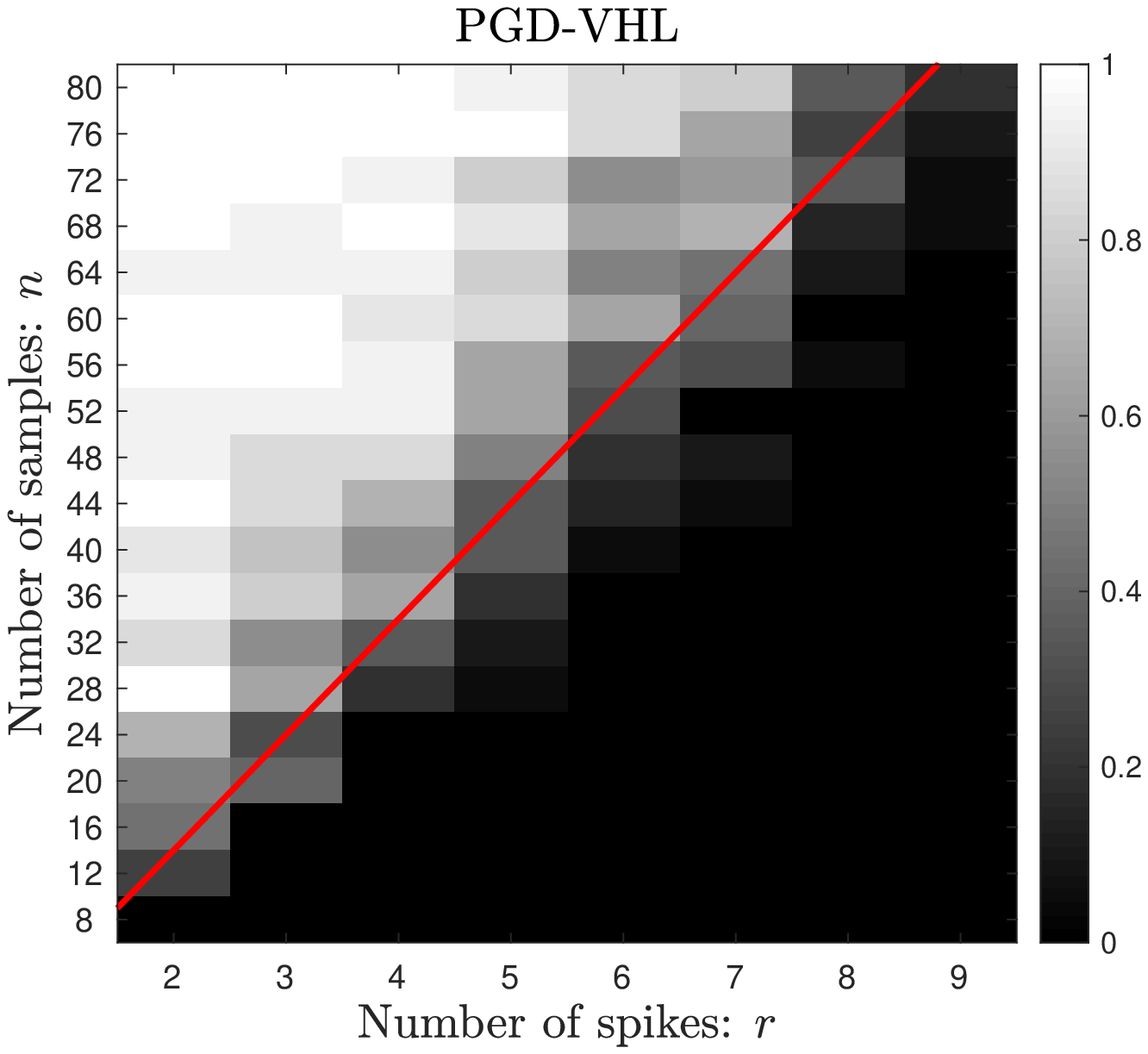}}
		%  \vspace{1.5cm}
		\centerline{(b)}\medskip
	\end{minipage}
	\caption{(a) The phase transition of PGD-VHL for varying $n$ and $s$ when $r = 4$. The red line plots the straight line $n = 2.5s$. (b) The phase transition of PGD-VHL for varying $n$ and $r$ when $s = 4$. The red line plots the straight line $n = 2.5r$.}
	\label{fig: phase transition n}
\end{figure}

\begin{figure}[htb]

	\begin{minipage}[b]{1\linewidth}
		\centering
		\centerline{\includegraphics[width=4.2cm]{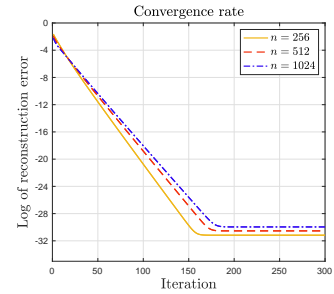}
	}
		%  \vspace{1.5cm}
	\end{minipage}
	\hfill
	
	\caption{Convergence of PGD-VHL for $n=256, 512, 1024$. Here we fix $s=4$ and $r=4$. }
	\label{fig: convergence}
\end{figure}

In the third simulation, we investigate the convergence rate of PGD-VHL for different $n$. The results are shown in Figure~\ref{fig: convergence}. The $y$-axis  denotes $\log\left(\fronorm{\mX_t - \mX^\natural}/\fronorm{\mX^\natural}\right)$ 
and the $x$-axis represents the iteration number. It can be clearly seen that PGD-VHL converges linearly as shown in our main theorem. %It is worth pointing out that PGD-VHL can be implemented under high dimensional regime, where $n=256,512,1024$. In these regimes, convex methods, like ANM and VHL, are unavailable to solve the problem.

\section{Conclusion}
\label{sec: conclusion}
In this paper, we propose an efficient algorithm named PGD-VHL towards recovering low rank matrix in blind super-resolution. Our theoretical analysis shows that the proposed algorithm converges to the target matrix linearly. This is also demonstrated by our numerical simulations.

\clearpage
     
\bibliographystyle{IEEEbib}
%\bibliography{strings,refs}

%\bibliographystyle{IEEEtranN}
\bibliography{refs}

\begin{thebibliography}{10}

\bibitem{quirin2012optimal}
Sean Quirin, Sri Rama~Prasanna Pavani, and Rafael Piestun,
\newblock ``Optimal 3{D} single-molecule localization for superresolution
  microscopy with aberrations and engineered point spread functions,''
\newblock {\em Proceedings of the National Academy of Sciences}, vol. 109, no.
  3, pp. 675--679, 2012.

\bibitem{qu2015accelerated}
Xiaobo Qu, Maxim Mayzel, Jian-Feng Cai, Zhong Chen, and Vladislav Orekhov,
\newblock ``Accelerated {NMR} spectroscopy with low-rank reconstruction,''
\newblock {\em Angewandte Chemie International Edition}, vol. 54, no. 3, pp.
  852--854, 2015.

\bibitem{luo2006low}
Xiliang Luo and Georgios~B Giannakis,
\newblock ``Low-complexity blind synchronization and demodulation for (ultra-)
  wideband multi-user ad hoc access,''
\newblock {\em IEEE Transactions on Wireless communications}, vol. 5, no. 7,
  pp. 1930--1941, 2006.

\bibitem{chi2016guaranteed}
Yuejie Chi,
\newblock ``Guaranteed blind sparse spikes deconvolution via lifting and convex
  optimization,''
\newblock {\em IEEE Journal of Selected Topics in Signal Processing}, vol. 10,
  no. 4, pp. 782--794, 2016.

\bibitem{yang2016super}
Dehui Yang, Gongguo Tang, and Michael~B Wakin,
\newblock ``Super-resolution of complex exponentials from modulations with
  unknown waveforms,''
\newblock {\em IEEE Transactions on Information Theory}, vol. 62, no. 10, pp.
  5809--5830, 2016.

\bibitem{li2019atomic}
Shuang Li, Michael~B Wakin, and Gongguo Tang,
\newblock ``Atomic norm denoising for complex exponentials with unknown
  waveform modulations,''
\newblock {\em IEEE Transactions on Information Theory}, vol. 66, no. 6, pp.
  3893--3913, 2019.

\bibitem{suliman2021mathematical}
Mohamed~A Suliman and Wei Dai,
\newblock ``Mathematical theory of atomic norm denoising in blind
  two-dimensional super-resolution,''
\newblock {\em IEEE Transactions on Signal Processing}, vol. 69, pp.
  1681--1696, 2021.

\bibitem{chen2020vectorized}
Jinchi Chen, Weiguo Gao, Sihan Mao, and Ke~Wei,
\newblock ``Vectorized hankel lift: A convex approach for blind
  super-resolution of point sources,''
\newblock {\em arXiv preprint arXiv:2008.05092}, 2020.

\bibitem{ahmed2013blind}
Ali Ahmed, Benjamin Recht, and Justin Romberg,
\newblock ``Blind deconvolution using convex programming,''
\newblock {\em IEEE Transactions on Information Theory}, vol. 60, no. 3, pp.
  1711--1732, 2013.

\bibitem{candes2009exact}
Emmanuel~J Cand{\`e}s and Benjamin Recht,
\newblock ``Exact matrix completion via convex optimization,''
\newblock {\em Foundations of Computational Mathematics}, vol. 9, no. 6, pp.
  717, 2009.

\bibitem{zhang2018multichannel}
Shuai Zhang, Yingshuai Hao, Meng Wang, and Joe~H Chow,
\newblock ``Multichannel {H}ankel matrix completion through nonconvex
  optimization,''
\newblock {\em IEEE Journal of Selected Topics in Signal Processing}, vol. 12,
  no. 4, pp. 617--632, 2018.

\bibitem{tu2016low}
Stephen Tu, Ross Boczar, Max Simchowitz, Mahdi Soltanolkotabi, and Ben Recht,
\newblock ``Low-rank solutions of linear matrix equations via procrustes
  flow,''
\newblock in {\em International Conference on Machine Learning}. PMLR, 2016,
  pp. 964--973.

\bibitem{zheng2016convergence}
Qinqing Zheng and John Lafferty,
\newblock ``Convergence analysis for rectangular matrix completion using
  burer-monteiro factorization and gradient descent,''
\newblock {\em arXiv preprint arXiv:1605.07051}, 2016.

\bibitem{chi2019nonconvex}
Yuejie Chi, Yue~M Lu, and Yuxin Chen,
\newblock ``Nonconvex optimization meets low-rank matrix factorization: An
  overview,''
\newblock {\em IEEE Transactions on Signal Processing}, vol. 67, no. 20, pp.
  5239--5269, 2019.

\bibitem{cai2018spectral}
Jian-Feng Cai, Tianming Wang, and Ke~Wei,
\newblock ``Spectral compressed sensing via projected gradient descent,''
\newblock {\em SIAM Journal on Optimization}, vol. 28, no. 3, pp. 2625--2653,
  2018.

\bibitem{cai2019fast}
Jian-Feng Cai, Tianming Wang, and Ke~Wei,
\newblock ``Fast and provable algorithms for spectrally sparse signal
  reconstruction via low-rank {H}ankel matrix completion,''
\newblock {\em Applied and Computational Harmonic Analysis}, vol. 46, no. 1,
  pp. 94--121, 2019.

\bibitem{suliman2018blind}
Mohamed~A Suliman and Wei Dai,
\newblock ``Blind two-dimensional super-resolution and its performance
  guarantee,''
\newblock {\em arXiv preprint arXiv:1811.02070}, 2018.

\bibitem{mao2021blind}
Sihan Mao and Jinchi Chen,
\newblock ``Fast blind super-resolution of point sources via projected gradient
  descent,''
\newblock {\em Under Preparation}, 2021.

\bibitem{cvx}
Michael Grant and Stephen Boyd,
\newblock ``{CVX}: Matlab software for disciplined convex programming, version
  2.1,'' Mar. 2014.

\end{thebibliography}

\end{document}